\documentclass[aps,pre,preprint]{revtex4-1}

\usepackage{graphicx}% Include figure files
\usepackage{subfig}
\usepackage{subfloat}
\usepackage{dcolumn}% Align table columns on decimal point
\usepackage{bm}% bold math
\usepackage[utf8]{inputenc}
\usepackage[T1]{fontenc}
\usepackage{mathptmx}
\usepackage{amsmath}
\usepackage{etoolbox}
\usepackage{hyperref}
\usepackage{xcolor}

\draft % marks overfull lines with a black rule on the right

\begin{document}

% Use the \preprint command to place your local institutional report number 
% on the title page in preprint mode.
% Multiple \preprint commands are allowed.
%\preprint{}

\title{Using test particle sum rules to improve approximations in classical DFT : White-Bear and White-Bear mark II versions of the Lutsko Functional.}  %Title of paper

% repeat the \author .. \affiliation  etc. as needed
% \email, \thanks, \homepage, \altaffiliation all apply to the current author.
% Explanatory text should go in the []'s, 
% actual e-mail address or url should go in the {}'s for \email and \homepage.
% Please use the appropriate macro for the type of information

% \affiliation command applies to all authors since the last \affiliation command. 
% The \affiliation command should follow the other information.
\author{Melih G\"ul}
\email[]{melih.guel@uni-tuebingen.de}
\author{Roland Roth}
\email[]{roland.roth@uni-tuebingen.de}
\affiliation{ Institute for Theoretical Physics, University of Tübingen, Auf der Morgenstelle 14, 72076 Tübingen, Germany}
\author{Robert Evans}
\email[]{bob.evans@bristol.ac.uk}
\affiliation{H. H. Wills Physics Laboratory, University of Bristol, Bristol BS8 ITL, United Kingdom}
%\homepage[]{Your web page}
%\thanks{}
%\altaffiliation{Institut f\"ur Theoretische Physik T\"ubingen}
%\affiliation{Eberhard-Karls-Universitaet Tuebingen}

\date{\today}

\begin{abstract}
In a recent paper [M. G\"ul et al., Phys. Rev. E, {\bf 110 (6)}, 064115] we showed that test particle sum rules, which address the excess chemical potential and isothermal compressibility, could be used to develop new and accurate classical density functionals for hard-sphere (HS) fluids. Here  we extend our approach to the construction of HS functionals building upon the state of art White-Bear (WB) and White-Bear mark II functionals. Employing the same test-particle sum rules we determine the two free parameters in the  Lutsko [James F. Lutsko, Phys. Rev. E, {\bf 102}, 062137] formulation of fundamental measure theory (FMT) by minimizing the relative errors between different routes to the two thermodynamic quantities. The resulting optimized Lutsko WB functionals, especially Lutsko WB mark II, are generally more accurate and consistent than those obtained in earlier treatments.
\end{abstract}

% insert suggested PACS numbers in braces on next line

\maketitle %\maketitle must follow title, authors, abstract and \pacs
\newpage
% Body of paper goes here. Use proper sectioning commands. 
% References should be done using the \cite, \ref, and \label commands
\section{Introduction}

It is well-known that sum rules play a key role across many-body physics. These are especially important in equilibrium classical statistical mechanics where they relate static structure, i.e. correlation functions, to bulk and interfacial thermodynamic properties. The papers of J.R. Henderson laid out most of the ideas and his 1992 review \cite{henderson1992}, titled succinctly 'Statistical Mechanical Sum Rules', remains the authoritative reference. \textcolor{black}{In our recent paper \cite{Melih2024} we investigated the efficacy of density functional theory (DFT) for one-component hard spheres (HS) by introducing two equilibrium sum rules, valid for any fluid described by a pair potential, that compare the excess chemical potential $\mu_{ex}$ and reduced isothermal compressibility $\chi_T$ calculated within the test-particle framework \cite{percus1962} with the corresponding bulk thermodynamic quantities.We chose to implement these general sum rules for the particular case of HS, in three dimensions, since i) this is arguably the most investigated system in the classical statistical mechanics of liquids , for which we have very accurate results for thermodynamics and structure,  and ii) HS are directly relevant and form a cornerstone in the physics of colloids. It is natural to treat hard spheres as a testing ground for developments in classical statistical physics.Specifically , improving upon existing HS functionals constitutes a challenge. In \cite{Melih2024} we optimized the two free parameters $A$ and $B$ entering the Lutsko functional \cite{Lutsko20} for HS by minimizing the differences between results obtained from the bulk thermodynamic equation of state and those from the test particle DFT route. Recall that in the latter the inhomogeneous one-body density profile, computed by fixing a particle at the origin so that the external potential is equal to the inter-particle pair potential, is proportional to the bulk radial distribution function \cite{Melih2024,percus1962}. Lutsko's paper had re-visited and extended the ideas of Rosenfeld et al. \cite{Rosenfeld97a,Rosenfeld97b} and suggested a means for constructing a new class of functionals in DFT \cite{Evans79, Hansen06} for HS within the framework of fundamental measure theory (FMT). In our paper\cite{Melih2024} we found that the Percus-Yevick (PY) approximation, corresponding to $8A+2B=9$, dictates the choices of parameters that perform most consistently across a wide range of HS packing fractions.}

The present paper continues to adopt Lutsko's approach for a HS functional but now we take a new stance by building the theory upon the White-Bear (WB) and White-Bear mark II (WBII) functionals \cite{Roth02,YuWu02,HansenGoos06}. By doing so, we aim to improve upon self-consistency with the sum rules for the excess chemical potential and isothermal compressibility \cite{Melih2024} and thereby construct more accurate functionals. Note that both WB functionals are known to be more accurate than the Rosenfeld (RF) version of FMT, which is constructed around the PY approximation, and both have found numerous applications. Both WB functionals are constructed around the accurate Carnahan-Starling (CS) bulk equation of state for HS. The difference between WB and WBII was expounded in \cite{Roth10}: the WB functional is inconsistent with respect to the pressure, i.e. there is a (slight) difference between the inputted CS pressure \cite{MCSL03, CS69} and the pressure obtained through the scaled-particle relation, required in the construction of FMT. In the WBII version, this inconsistency is remedied \cite{HansenGoos06}. Here we follow the same methodology as in \cite{Melih2024} in order to determine optimized values for the  Lutsko parameters $A$ and $B$ for the new Lutsko White -Bear functionals. The same combination $8A+2B=9$ now corresponds to CS for both versions (LK-WB and LK-WBII) of the Lutsko functional.

\textcolor{black} {Our paper is arranged as follows: in Sec.II we remind readers of the formalism presented in \cite{Melih2024} and explain how we modify the Lutsko HS functional to incorporate the WB functionals. Section III describes the results of our calculations using the new functionals and how we obtain optimized values for parameters $A$ and $B$. This Section also describes how well our new (optimized) functionals perform in the calculation of the pressure, virial coefficients, the bulk pair direct correlation function $c^{(2)}(r)$ and provides an important example of density profiles corresponding to confinement of HS in a small hard spherical cavity.We summarize in Sec.IV.}

\section{Extending the Lutsko HS functional} \label{sec:lutkso-functional}

In our previous paper \cite{Melih2024}, we investigated the Lutsko functional
\begin{equation}\label{eq::lutsko-rf}
    \Phi^\text{LK}_\text{ex} = \Phi_1+\Phi_2+\Phi_3^\text{LK},
\end{equation}
where $\Phi_1$ and $\Phi_2$ are the first two parts of the standard Rosenfeld (RF) functional \cite{Rosenfeld89}. The third part $\Phi_3^\text{LK}$ is
\begin{equation}\label{eq::lutsko-rf-part3}
    \Phi_3^\text{LK} = f^\text{RF}(n_3)\left[(A+B)n_2^3-3 A~ n_2 \mathbf{n}_2\cdot\mathbf{n}_2+3 A~ \mathbf{n}_2~\mathbf{T}~\mathbf{n}_2-3 B~n_2\text{Tr}\,\mathbf{T}^2+(2 B-A)\text{Tr}\,\mathbf{T}^3\right],
\end{equation}
where
\begin{equation}\label{eq::f-rf}
    f^\text{RF}(n_3)=\frac{1}{24\pi(1-n_3)^2},
\end{equation}
can be regarded as a generalization of the tensorial version of the RF functional \cite{Tarazona2000}. The weighted densities $n_\alpha(\textbf{r})$ for a one-component system are defined as 
\begin{equation}\label{eq::wds}
    n_\alpha(\textbf{r})=\int \text{d}\mathbf{r}'\,\rho(\textbf{r}') w_\alpha(\textbf{r}-\textbf{r}'),
\end{equation}
where $\rho(\mathbf{r})$ is the density profile and $w_\alpha(\textbf{r})$ are the weight functions of FMT \cite{Rosenfeld89}. A generalization of Eq.\eqref{eq::wds} to multi-component hard-sphere mixture is straightforward. Specifically for the tensorial weighted density $\textbf{T}(\textbf{r})$, we have
\begin{equation}\label{eq::tensor-wd}
    \textbf{T}(\textbf{r}) = \int \text{d}\textbf{r}'\,\rho(\textbf{r}')w_T(\textbf{r}-\textbf{r}')
\end{equation}
with the tensor weight function
\begin{equation}\label{eq::t-weight-function}
    w_T(\textbf{r}) = \textbf{e}_r\otimes\textbf{e}_r\,\delta(R-|\textbf{r}|)
\end{equation}
where $\mathbf{e}_r=\mathbf{r}/r$ is the radial unit vector.
For $A=-B=3/2$ the Lutsko functional recovers the tensor version of the Rosenfeld functional. The parameters $A$ and $B$ were introduced by Lutsko \cite{Lutsko20} when he revisited the ideas of dimensional crossover. These additional parameters offer the possibility to tune the corresponding functional in such a way that certain thermodynamic and structural properties are improved. We employed sum rules for the excess chemical potential $\mu_\text{ex}$ and (reduced) isothermal compressibility $\chi_T$ \cite{Melih2024}. Both thermodynamic quantities can be obtained from a) the bulk free energy density $\Phi$ of the corresponding functional and b) employing test particle geometry \cite{Melih2024} where the excess chemical potential is given as the change of grand potential, $\Delta\Omega$, due to insertion of a solute test particle identical to the solvent particles.
Our paper \cite{Melih2024} laid out how $\Delta\Omega$ is calculated within DFT. Furthermore, the isothermal compressibility can be obtained noting that the structure factor $S(k=0)$, at vanishing wave number, is equal to the reduced compressibility$\chi_T$ \cite{Hansen06}
\begin{equation}\label{eq::xt}
    S(k=0) = 1+4\pi\rho_b\int_0^\infty \text{d}r\,r^2 (g(r)-1) = \chi_T,
\end{equation}
where $g(r)$ is the radial distribution function of the uniform fluid of density $\rho_b$. Note that $\chi_T=\rho_bk_B T\kappa_T$, and $\kappa_T$ is the usual bulk isothermal compressibility. The associated computation in this second route requires the corresponding density profile $\rho(r;\phi)$ in test particle geometry that we obtain by minimizing the grand functional $\Omega[\rho(r)]$ with a particle fixed at the origin exerting an external potential identical to the pair potential $\phi(r)$ : $\rho(r;\phi)=\rho_b g(r)$. It follows that the corresponding excess functional ${\cal F}_\text{ex}[\rho]$ can be interrogated  with respect to these two test particle sum rules, providing a way to determine meaningfully the Lutsko parameters $A$ and $B$. Sec.(IIA) in \cite{Melih2024} outlines the sum rules and this strategy in more detail.

Recalling that the White-Bear (WB) functionals are somewhat more accurate and self-consistent than the RF functional, it is compelling to extend the form of the Lutsko functional, Eq.\eqref{eq::lutsko-rf}, such that the tensorial version of WB and WB mark II (WBII) can be recovered. We might expect a higher degree of self-consistency for these types of functionals. As the WB functional differs merely in the $n_3$-dependence of $\Phi_3$ in the RF functional, it is appealing to apply the same modification also to $\Phi^\text{LK}_3$, Eq.\eqref{eq::lutsko-rf-part3}, of the Lutsko functional. The same approach will also apply to WBII.

In order to introduce the WB-Lutsko functional, we build upon the FMT-toolbox idea \cite{Roth10} that highlights the different roles of the functions of $n_3$ and of the other weighted densities. To this end we write 
\begin{equation}\label{eq::lutsko-wb}
	\Phi_{3}^{\text{LK-WB}}=f(n_3)\left[(A+B)n_2^3-3 A~ n_2 \mathbf{n}_2\cdot\mathbf{n}_2+3 A~ \mathbf{n}_2~\mathbf{T}~\mathbf{n}_2-3 B~n_2\text{Tr}\,\mathbf{T}^2+(2 B-A)\text{Tr}\,\mathbf{T}^3\right],
\end{equation}
with
\begin{equation}\label{eq::wb-n3}
	f(n_3)=\frac{n_3+(1-n_3)^2\log(1-n_3)}{36\pi n_3^2(1-n_3)^2},
\end{equation}
which replaces the corresponding expression $f^\text{RF}(n_3)$ of the Rosenfeld functional. 
In total the Lutsko-WB functional is:
\begin{equation}\label{eq::lutsko-wb}
    \Phi^{\text{LK-WB}}_\text{ex}=\Phi_1+\Phi_2+\Phi_3^{\text{LK-WB}}.
\end{equation}
The WBII version of the Lutsko functional is somewhat more complex:
\begin{equation}\label{eq::lutsko-wbII}
	\Phi^{\text{LK-WBII}}_\text{ex}=\Phi_1+\varphi_1(n_3)\Phi_2+\varphi_2(n_3)\Phi_3^{\text{LK}},
\end{equation}
where
\begin{align}\label{eq::wbII-phis}
	\varphi_1(n_3) &= 1+\frac{2n_3-n_3^2+2(1-n_3)\log(1-n_3)}{3n_3}\\\nonumber
	\varphi_2(n_3) &= 1-\frac{2n_3-3n_3^2+2n_3^3+2(1-n_3)^2\log(1-n_3)}{3n_3^2}.
\end{align}
are the functions entering the original derivation in the WBII construction \cite{HansenGoos06}.

Thus, we generalize the Lutsko functional to its WB and WBII versions by replacing the numerator of the corresponding $\Phi_3$ of WB or WBII with the numerator in square brackets given in Lutsko's functional, Eq.\eqref{eq::lutsko-rf-part3}, containing the parameters $A$ and $B$ \cite{Lutsko20}. By construction, these new functionals recover the tensorial versions of WB and WBII for $A=-B=3/2$ and therefore the bulk thermodynamic quantities of CS.
It will be convenient to abbreviate the Lutsko-WB functional as LK-WB$(A,B)$ and analogously LK-WBII$(A,B)$ for the mark II version.

The corresponding equation of states (EoS) for the uniform one-component hard-sphere fluid of density $\rho$ are obtained from the bulk thermodynamic route $\beta P=-\Phi+\rho\partial\Phi/\partial\rho$. Here, $\Phi(\rho)=\Phi_\text{ex}(\rho)+\Phi_\text{id}(\rho)$ is the bulk Helmholtz free energy density with the ideal contribution $\Phi_\text{id}=\rho(\log(\Lambda^3\rho)-1)$ where $\Lambda$ is the thermal wavelength. Hence, we find
\begin{equation}\label{eq::Lutsko-wb-eos}
	\frac{\beta P^{\text{LK-WB}}(A,B)}{\rho}=\frac{\beta P^{\text{CS}}}{\rho}+(8A+2B-9)\frac{\eta^2(3-\eta)}{9(1-\eta)^3}
\end{equation}
and
\begin{equation}\label{eq::Lutsko-wbII-eos}
	\frac{\beta P^{\text{LK-WBII}}(A,B)}{\rho}=\frac{\beta P^{\text{CS}}}{\rho}+(8A+2B-9)\frac{\eta^2(3-2\eta+\eta^2)}{9(1-\eta)^3},	
\end{equation}
where $P^{\text{CS}}$ is the Carnahan-Starling (CS) pressure, $\eta=\rho\pi\sigma^3/6$ is the packing fraction and $\sigma$ the HS diameter. Again, we see that for $8A+2B=9$ both EoS, Eq.\eqref{eq::Lutsko-wb-eos} and \eqref{eq::Lutsko-wbII-eos}, reduce to the CS EoS. Note that $P^{\text{LK-WBII}}$ is modified differently from $P^{\text{LK-WB}}$.
It is important to note that the self-consistency of the LK-WB functional with respect to the pressure $P^\text{SP}$ from the scaled-particle route
\begin{equation}\label{eq::pressure-consistency}
    \beta P^{\text{SP}} = \frac{\partial\Phi}{\partial n_3},
\end{equation}
is violated, in accordance with the original discussion, given in \cite{Roth02}. The associated deviation between the pressure from Eq.\eqref{eq::Lutsko-wb-eos} and the scaled-particle (SP) pressure obtained from Eq.\eqref{eq::pressure-consistency} will now depend on the Lutsko parameters $A$ and $B$. This implies that full consistency of the pressure can be realized by choosing the Lutsko parameters accordingly. Due to its construction, the scaled-particle result, Eq.\eqref{eq::pressure-consistency}, is automatically fulfilled in the case of LK-WBII, provided $8A+2B=9$. Note that in the case of the Lutsko functional introduced in \cite{Melih2024} the self-consistency is automatically fulfilled. The reason lies in the fact that the $n_3$-dependence of the Lutsko Helmholtz free energy density, see Eq.\eqref{eq::lutsko-rf-part3}, is the same as that of the original Rosenfeld functional which, from its derivation, returns the same expression for the pressure when calculated via the thermodynamic route and the scaled-particle route.

\section{Results}\label{results}

The numerical methods that we employ are identical to those discussed in Sec.III of \cite{Melih2024} and the same analysis is employed, i.e. we optimize the Lutsko-WB functionals Eq.\eqref{eq::lutsko-wb} and Eq.\eqref{eq::lutsko-wbII} by minimizing the relative error between the bulk and DFT route for the excess chemical potential $\mu_\text{ex}$ and for the reduced isothermal compressibility $\chi_T$:
\begin{equation}\label{eq::rel-deviations}
    \delta_\mu = \frac{\mu^\text{DFT}_\text{ex}-\mu^\text{Bulk}_\text{ex}}{\mu^\text{Bulk}_\text{ex}},\quad \delta_\chi=\frac{\chi_T^\text{DFT}-\chi_T^\text{Bulk}}{\chi_T^\text{Bulk}} 
\end{equation}
with respect to the parameters $A$ and $B$ hence providing an optimal point $(A,B)$ of the corresponding functional. We chose to minimize $M(\delta_\mu, \delta_\chi)=\frac{1}{2}(|\delta_\mu|+|\delta_\chi|)$; in principle any other suitable function would be valid \cite{Melih2024}. Superscript Bulk in Eq.\eqref{eq::rel-deviations} refers to results obtained from the bulk free energy density $\Phi(\rho)$, as used to obtain Eq.\eqref{eq::Lutsko-wb-eos} and Eq.\eqref{eq::Lutsko-wbII-eos}. Explicit expressions are given in Appendix  \ref{app::thermoQuant}. We compute $\mu^\text{DFT}_\text{ex}$ and $\chi^\text{DFT}_T$ as outlined in Sec.\ref{sec:lutkso-functional}, i.e. in the test-particle geometry with the spherically symmetric equilibrium density profile $\rho(r;\phi)$ obtained from the corresponding minimization of the grand potential functional.

We emphasize that our analysis takes into account only high densities where considerable deviations occur, see also \cite{Melih2024}. This, of course, assumes that the optimal points will lie close to what we term the CS-line $8A+2B-9=0$ so that for low densities the associated deviations will remain small. This argument follows directly from the fact that the functionals are exact up to second order in density therefore yielding small errors in the thermodynamic quantities at low densities. 
In practice we calculate an average relative deviation for $\delta_\mu$ and $\delta_\chi$ using three high densities in order to derive the optimal point, i.e. we consider packing fractions $\eta=0.35, 0.40$ and $0.45$, weighted equally. This approach ensures that the corresponding optimal point accounts for  states throughout the high-density regime. 

With the same ranges of $A$ and $B$ as given in Sec.(IV) in \cite{Melih2024}, the following optimal points are found
\begin{align}\label{eq::optimal-points}
    \text{LK-WB}:&\quad A=1.35,\quad B=-0.85\\\nonumber
    \text{LK-WBII}:&\quad A=1.25,\quad B=-0.2,
\end{align}

\begin{figure}[h]
\includegraphics[width=0.49\linewidth]{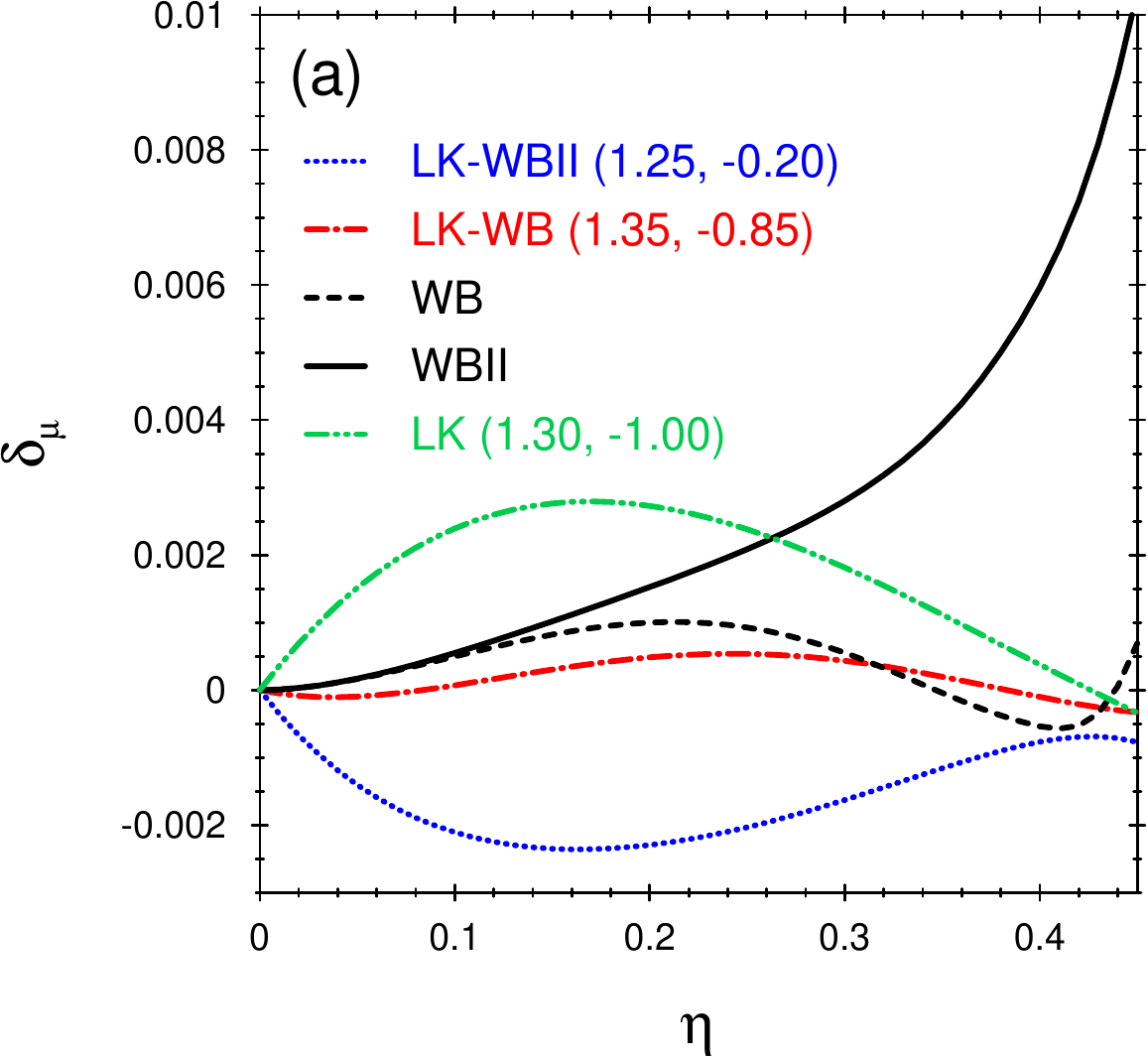}
\includegraphics[width=0.49\linewidth]{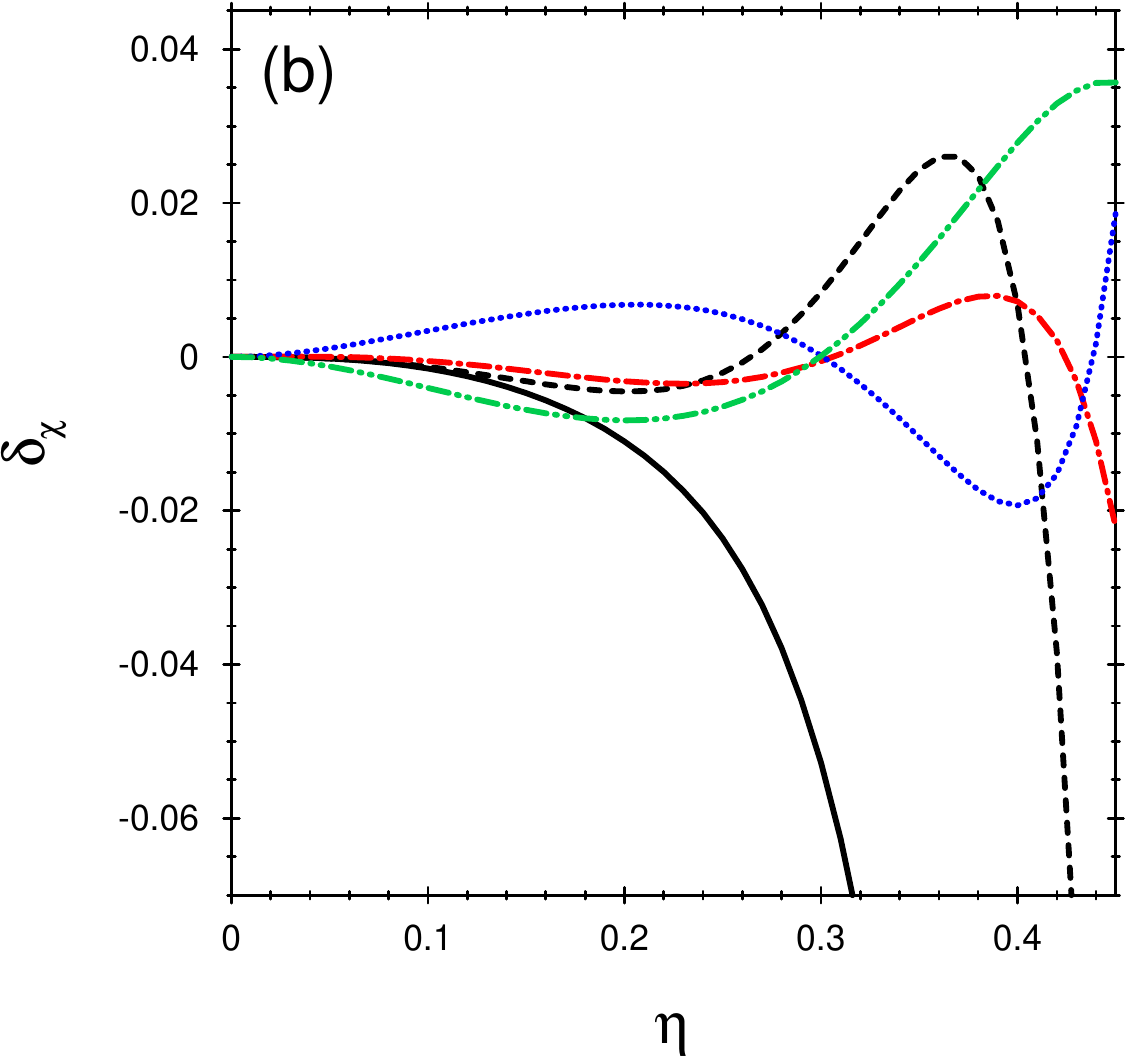}
\caption{\label{fig:sumrule_devs_mu_xt} Relative deviations (a) $\delta_\mu$ and (b) $\delta_\chi$ vs.the packing fraction $\eta$. While the original versions of the White-Bear (WB) (dashed black) and White-Bear mark II (WBII) (black) functionals do not possess adjustable parameters, the Lutsko versions of these HS functionals (red and blue, respectively) allow us to minimize the relative deviations by adjusting the parameters $A$ and $B$, as explained in the text. We also show our previous result for the optimized Lutsko-Rosenfeld functional (green).}
\end{figure}
The LK-WB result is quite close to LK(1.3, -1.0) found in \cite{Melih2024}.
Moreover ($8A+2B-9$\,=\,0.1) is close to the CS-line whereas LK-WBII is much more distant ($8A+2B-9$\,=\,0.6). Similar to Figs.(2) and (3) of \cite{Melih2024}, the lines of smallest deviations of the aforementioned sum rules, Eq.\eqref{eq::rel-deviations}, lie close to the CS-line.

\begin{figure}[h]
\includegraphics[width=0.49\linewidth]{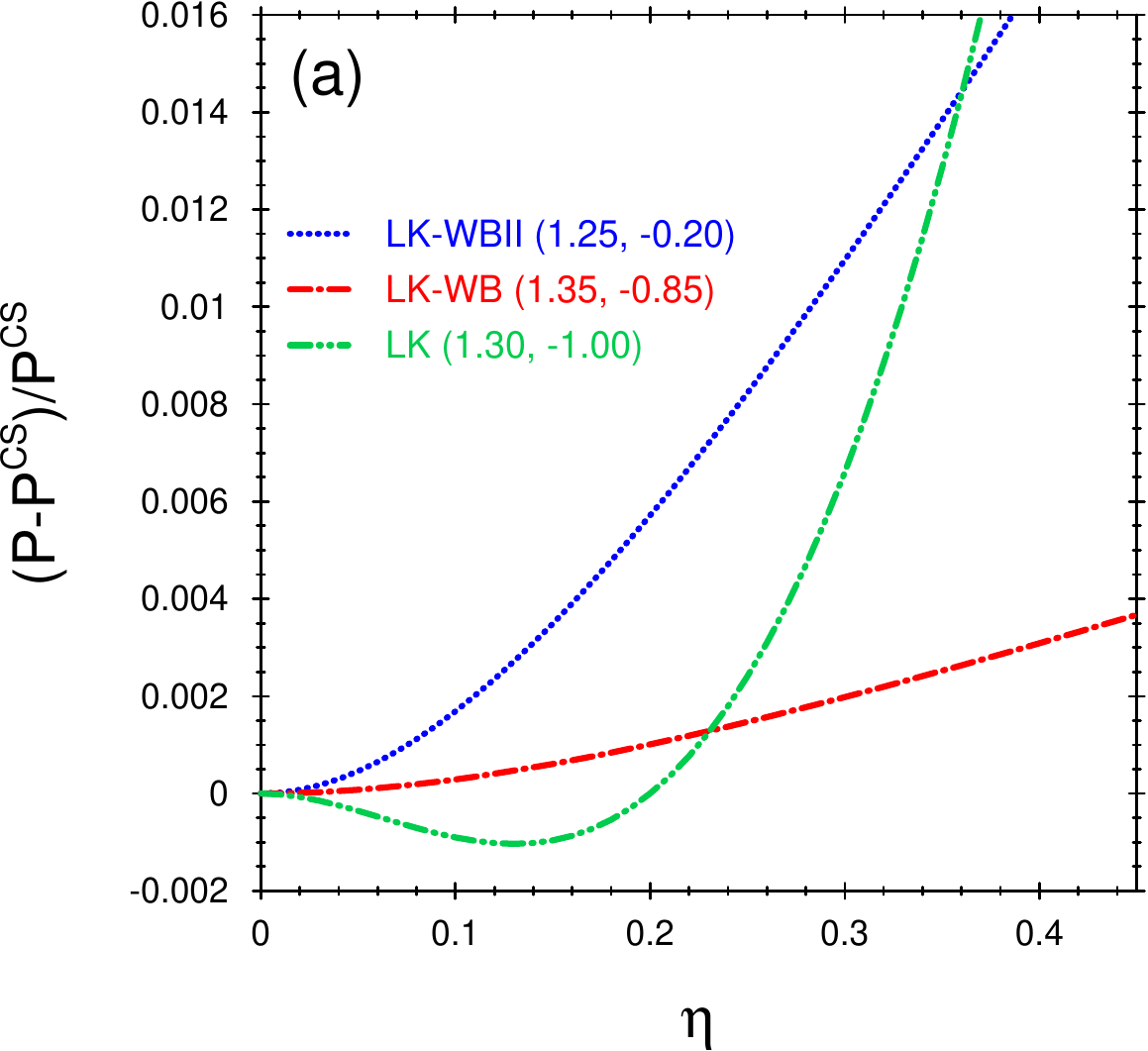}
\includegraphics[width=0.49\linewidth]{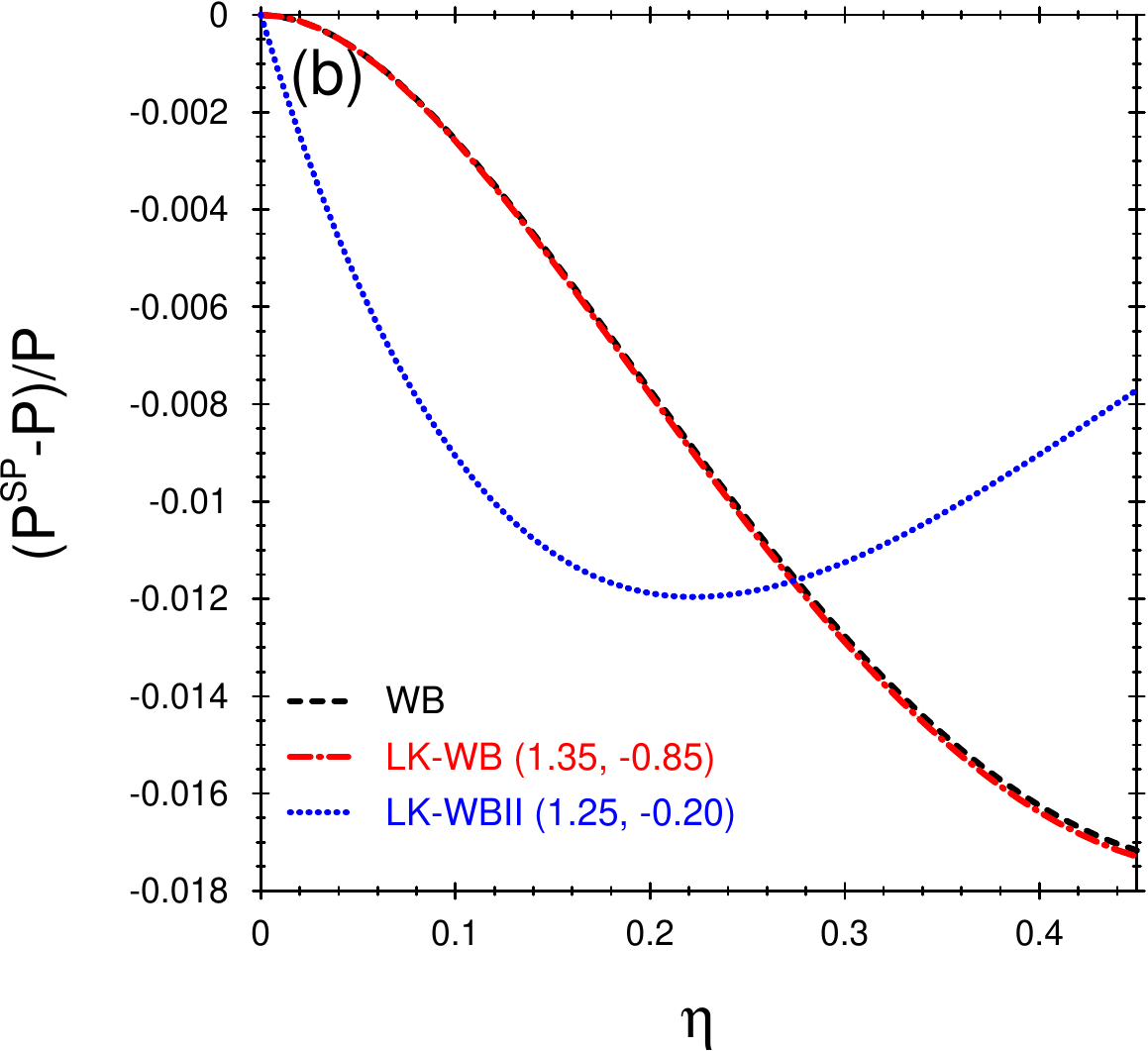}
\caption{\label{fig:pressures} (a) displays relative deviations of the HS pressures obtained from the Lutsko functionals with respect to the CS pressure $P^{\text{CS}}$. (b) shows the deviation $(P^{\text{SP}}-P)/P$ between $\beta P$ obtained via Eq.\eqref{eq::Lutsko-wb-eos}, \eqref{eq::Lutsko-wbII-eos} and the scaled-particle result $\beta P^{\text{SP}}=\partial \Phi/\partial n_3$ for the two LK-WB functionals and the original WB functional. We see that deviations remain below 2\%, showing the high degree self-consistency.}
\end{figure}

\begin{figure}
\includegraphics[width=0.6\linewidth]{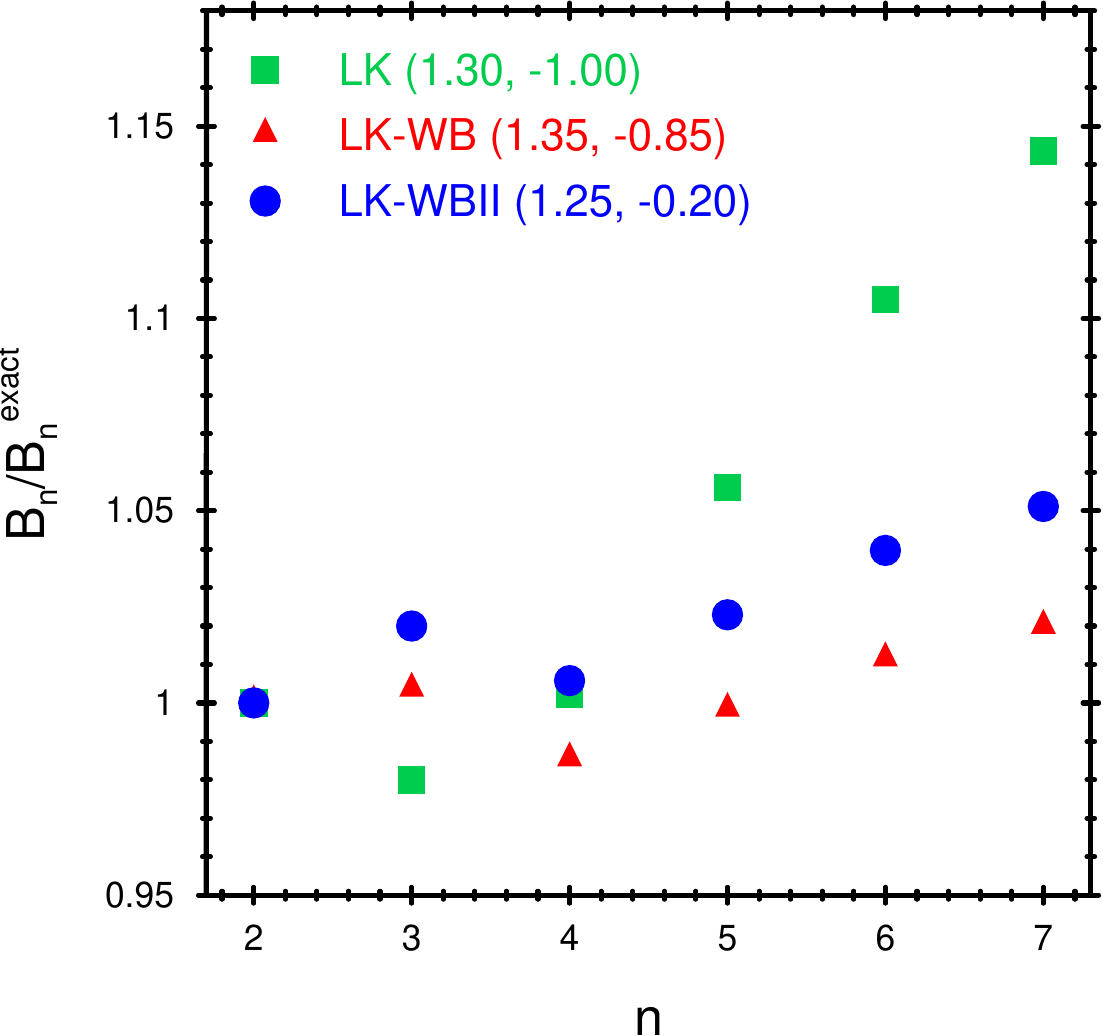}
\caption{Virial coefficients $B_n$ of the optimized Lutsko HS functionals normalized to the exact values $B_n^\text{exact}$ taken from \cite{Labik2005}.\label{fig::virials}}
\end{figure}

The relative deviations $\delta_\mu$ and $\delta_\chi$ defined in Eq.\eqref{eq::rel-deviations} and in \cite{Melih2024}, are presented as a function of the packing fraction $\eta$ in Fig.~\ref{fig:sumrule_devs_mu_xt}. For comparison, we again display the curves (green) of LK$(1.3,-1.0)$ obtained in \cite{Melih2024}. \textcolor{black} {It is interesting to observe that the magnitude of the relative deviations in the case of WBII, considered to be the most accurate HS functional, increase rapidly, well above the other results, for $\eta>0.35$.} We also infer for lower packing fractions that LK$(1.3,-1.0)$ and LK-WBII$(1.25,-0.20)$ already have high relative deviations since they are most distant from the PY- and CS-line, respectively. This becomes clear when we consider the leading virial coefficients which already start to deviate from the exact found ones, see Fig.~\ref{fig::virials}. For LK-WB$(1.35,-0.85)$ we find that the relative deviations remain small over the whole range of $\eta$ we consider here. Specifically for $\eta>0.35$ the relative deviations are smaller than those of WB, and smaller than LK$(1.30,-1.00)$ and LK-WBII$(1.25,-0.20)$. Noting the proximity to the CS-line, the relative deviations of LK-WB$(1.35,-0.85)$ are small for low densities and increase slowly for higher densities. We note that the excess chemical potential $\mu_\text{ex}$ is very accurately described by LK-WB$(1.35,-0.85)$, across the full range of $\eta$.

Overall this test of the sum rules for $\mu_{ex}$ and $\chi_T$ suggests we can improve upon the accuracy of LK$(1.3,-1.0)$ for high densities. Particularly for the case of WBII, introducing its Lutsko version, Eq.\eqref{eq::lutsko-wbII}, significantly improves consistency with the sum rules. However, this is less accurate than LK-WB$(1.35,-0.85)$ considering the whole range of liquid packing fractions $\eta$.

Figure~\ref{fig:pressures}(a) shows the relative deviations of the Lutsko functional predictions with respect to the CS pressure. Whilst LK$(1.30,-1.00)$ and LK-WBII$(1.25,-0.20)$ deviate from $P^{\text{CS}}$ for high densities, LK-WB$(1.35,-0.85)$ stays close; this is expected as the latter is close to the CS-line. Note that in contrast to WBII, which is constructed to yield the CS pressure for the one-component HS fluid, LK-WBII$(1.25,-0.20)$ exhibits deviations; the optimized Lutsko parameters do not satisfy $8A+2B-9=0$.

In Figure~\ref{fig:pressures}(b) we plot deviations between the pressure computed via the bulk thermodynamic route and via the relation $\partial\Phi/\partial n_3$ from scaled-particle theory (SP) \cite{SPT04}. We see that the deviations for WB and LK-WB$(1.35,-0.85)$ are almost the identical. These increase for high densities but still lie below 2\%. LK-WBII$(1.25,-0.20)$ shows deviations of similar magnitude but the variation with $\eta$ is different. Recall that while WBII guarantees consistency between routes, LK-WBII$(A,B)$ is inconsistent if $8A+2B-9\neq 0$, as is the case here. We infer that LK-WBII$(1.25,-0.20)$ maintains consistency to better than 1.2\% across the range of $\eta$.

The virial coefficients $B_n$ of the various Lutsko functionals are derived from the corresponding EoS by expanding with respect to density, which, for the optimized parameters, are shown in Fig.~\ref{fig::virials} for $n=2,\dots,7$. The Lutsko functionals are exact up to second order, but start to deviate from the exact result for $n>2$. LK-WB$(1.35,-0.85)$ and LK-WBII$(1.25,-0.20)$ virial coefficients stay close to the exact values, the former being slightly more accurate, especially for higher orders. This is consistent with the fact that the optimal point of LK-WB is close to the CS-line. By contrast, LK$(1.30,-1.00)$ starts to overestimate significantly the exact values already for $n\geq5$.
Tab.\ref{tab:virials} lists the virial coefficients of the Lutsko functionals up to $n=7$.

\begin{figure}
\includegraphics[width=0.6\linewidth]{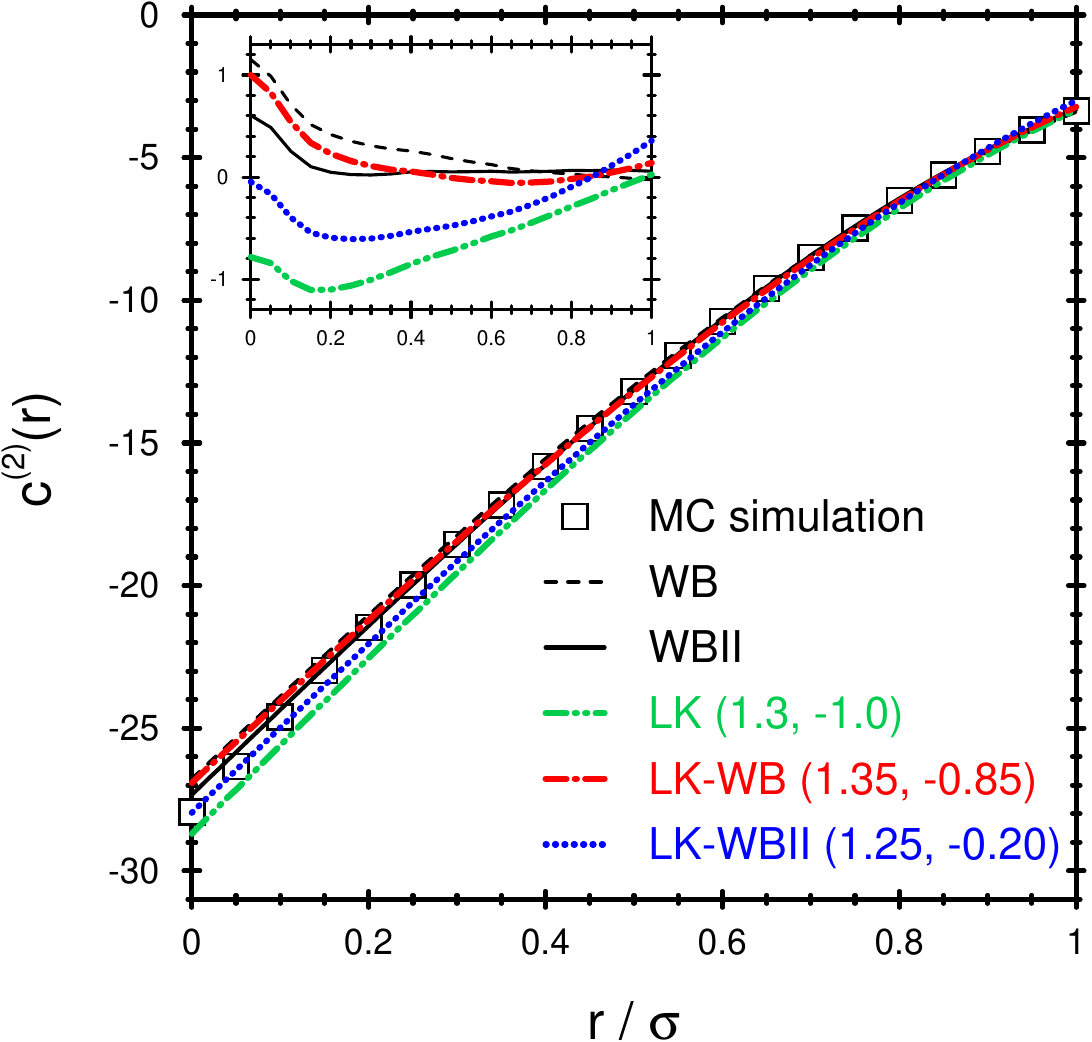}
\caption{\label{fig:c2s} The bulk pair direct correlation function $c^{(2)}(r)$ of HS at a packing fraction $\eta=0.419$ for WB, WBII, LK$(1.3,-1.0)$, LK-WB$(1.35,-0.85)$ and LK-WBII$(1.25,-0.20)$ together with MC simulation data taken from \cite{Groot87a}. \textcolor{black}{The inset shows the difference $c^{(2)}(r)-c^{(2)}_\text{MC}(r)$ between results for the respective functional and the MC data}.}
\end{figure}

As in \cite{Melih2024}, we also consider the pair direct correlation functions $c^{(2)}(r)$ resulting from the functionals of interest for packing fraction $\eta=0.419$, see Fig.~\ref{fig:c2s}.\textcolor{black} { Recall early treatments of DFT placed much emphasis on generating accurate  $c^{(2)}(r)$. For $r\approx 0$ we see that the WB, WBII and LK-WB$(1.35,-0.85)$ results are slightly above the Monte Carlo ( MC ) simulation values whereas LK$(1.3,-1.0)$ and LK-WBII$(1.25,-0.20)$ results are slightly below. \textcolor{black}{The inset of Fig.~\ref{fig:c2s} displays the difference between the pair direct correlation function of the corresponding functional and the MC data. We observe that WBII (black) performs well over the whole range of $r<\sigma$, better than LK-WBII$(1.25,-0.20)$ (blue) , except very close to $r=0$ and we see an improvement of LK-WB$(1.35,-0.85)$ (red) compared to WB (black dashed.)} We also attempted to optimize the Lutsko parameters $A$ and $B$ using the direct pair correlation function at this fixed packing fraction $\eta=0.419$. First, keeping the values $c^{(2)}(r=0)$ and $c^{(2)}(r=\sigma)$ fixed to MC simulation data, we find for LK-WB $A=1.48$ and $B=-0.94$, which are close to the values found by optimizing with respect to the sum rules. For LK-WBII, we find $A=1.60$ and $B=-1.61$, values quite distant from the optimal values found via the sum rules. Second, we can minimize the mean-squared error between MC simulation data and the DFT result, yielding for LK-WB $A=1.73$ and $B=-2.12$ and for LK-WBII $A=1.72$ and $B=-2.21$, values that are again quite different from those obtained using the sum rules.Although all the functionals employed in Fig.~\ref{fig:c2s} provide rather accurate $c^{(2)}(r)$, certainly better than RF (equivalent to Percus-Yevick) (see Ref. 2.),the region near $r\approx 0$ depends sensitively on the choice of functional. The choice LK-WBII$(1.25,-0.20)$ performs well. However,as mentioned earlier, it is less accurate overall than WBII. We concluded that optimizing w.r.t. $c^{(2)}(r)$ is probably not a profitable way forward, at least for HS, and we did not pursue further fitting.}
\begin{figure}
\centering
\includegraphics[width=0.6\linewidth]{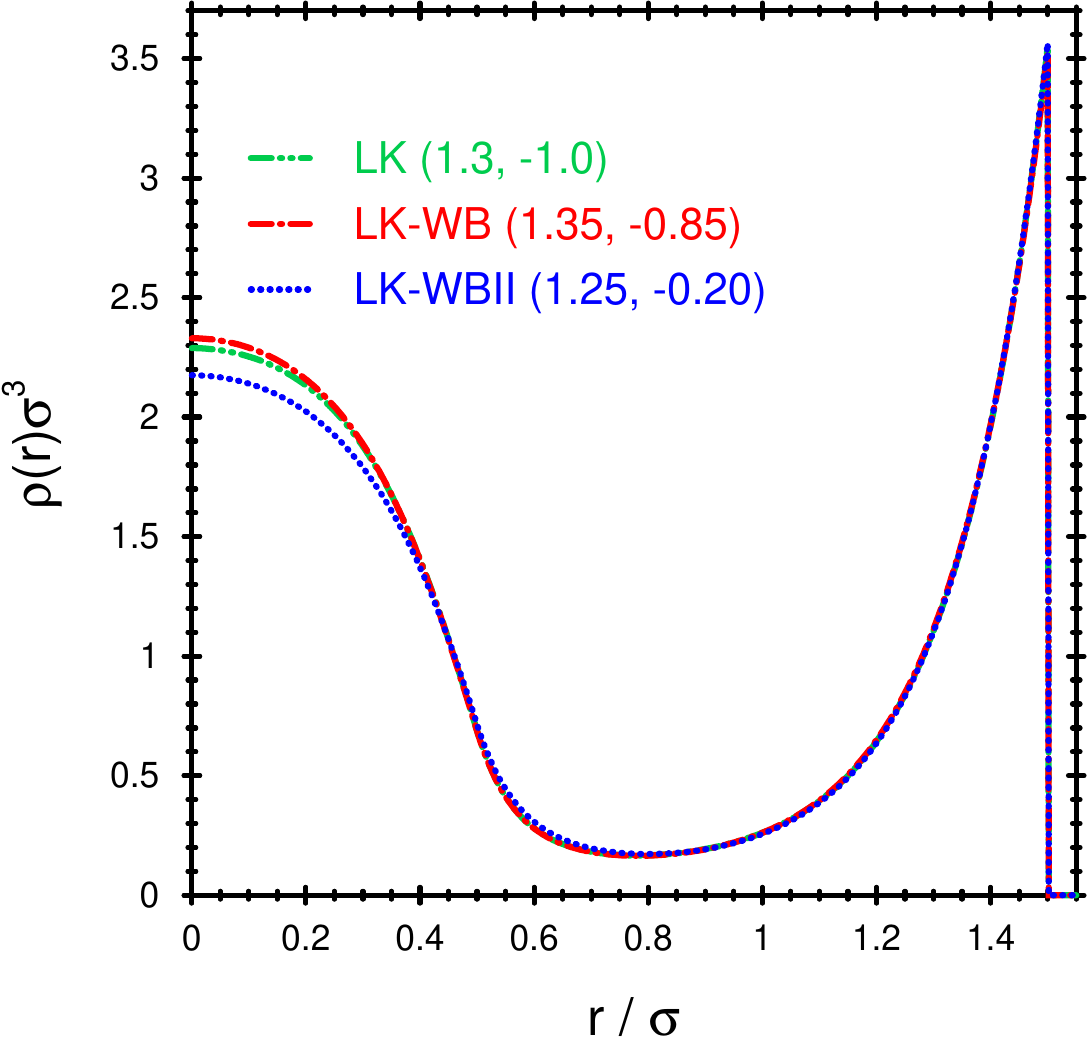}
\caption{\label{fig:sp-cav} Spherically symmetric density profiles $\rho(r)$ obtained using LK$(1.3, -1.0)$, LK-WB$(1.35, -0.85)$ and LK-WBII$(1.25, -0.20)$ for HS confined in a spherical cavity of radius $R=2\sigma$ with reservoir packing fraction $\eta=0.32$.}
\end{figure}

Finally, we report some observations on the stability of the Lutsko HS functionals that we consider here. In Fig.~\ref{fig:sp-cav} we present the density profiles based on the functionals LK$(1.3, -1.0)$, LK-WB$(1.35, -0.85)$ and LK-WBII$(1.25, -0.20)$ for the HS fluid confined in a hard spherical cavity of radius $R=2\sigma$, centered at the origin, and where the reservoir packing fraction $\eta=0.32$. This is a strongly confined situation reflected in the shape of the density profile; note the pronounced minimum around $r=0.8\sigma$ and the sharp peak at the contact value which corresponds to $r=3/2\sigma$ for this choice of cavity. Differences between results from the functionals are noticeable near the center of the spherical cavity but otherwise the profiles are near identical.
\textcolor{black}{Whilst these three  Lutsko functionals do not show any signs of instability when minimizing the grand potential functional for such a strongly confined system at relatively high packing fractions, the Rosenfeld (RF) and White-Bear (WB) functional fail in this test case. Specifically the minimization procedure does not converge to an equilibrium solution. We connect this shortcoming of RF and WB to the 0$D$-situation discussed in \cite{Rosenfeld96,Rosenfeld97a}, where potential divergent behavior of $\Phi_3$ is emphasized. Modifications of RF were introduced, including the anti-symmetrized version --or $q_3$-version for short -- that possesses the correct $0$D-limit and does not alter the thermodynamic properties of the bulk fluid. (Note that the original Rosenfeld papers  \cite{Rosenfeld89,Rosenfeld93} had already raised the problem of applying $\Phi_3$  to highly peaked density profiles in crystals.)  By comparing results of these modifications with those of the LK-WB and LK-WBII functionals given in Fig.~\ref{fig:sp-cav} , we observe only very minor differences. The $q_3$-version, together with other modifications, were also studied by Gonzalez.et.al. \cite{gonzalez1998} where results were compared to grand-canonical MC simulations of HS in several hard spherical cavities. These showed  only small differences between the density profiles at the center of the cavity - of a similar extent to those we observe in Fig.~\ref{fig:sp-cav}. The authors concluded that the quasi-0$D$ situation is well-captured by the aforementioned modifications of the RF functional. We draw the same conclusion here for the Lutsko functionals that we consider. It is interesting to recall that Lutsko showed \cite{Lutsko20} that explicit stability,  meaning the free energy is bounded from below, is guaranteed if $A,\,B$ are positive or zero whereas other cases remain unclear. The Lutsko functionals we employ clearly do not belong to the explicitly stable class. However, they yield stable equilibrium density profiles in this challenging case.}

\section{Summary}
We have investigated new functionals for the hard sphere fluid employing the formulation of FMT in Lutsko's paper \cite{Lutsko20} by determining his parameters \textit{A, B} using two equilibrium (test particle) sum rules, namely those for the excess chemical potential and isothermal compressibility. These were introduced and employed earlier \cite{Melih2024} in an analysis based upon the original Rosenfeld formulation \cite{Rosenfeld89} of FMT. The new versions, outlined in Sec.\ref{sec:lutkso-functional}, are based upon more accurate functionals for hard spheres, both employing the CS equation of state, namely White-Bear and White-Bear mark II \cite{Roth02,YuWu02,HansenGoos06}. It turns out that in the construction of these new functionals -- see Eq.\eqref{eq::lutsko-wb} and Eq.\eqref{eq::lutsko-wbII} -- the tensorial versions of WB and WBII are recovered for $8A+2B=9$, the same relation between parameters that was obtained in \cite{Melih2024}. The optimization of parameters follows that in \cite{Melih2024} and our results show that the new functionals perform somewhat better than earlier functionals. In particular LK-WBII(1.25,-0.20) improves significantly upon WBII regarding consistency with sum rules.

\textcolor{black} { Most of our paper focused on improvements of consistency to \textit{bulk }thermodynamics Figures 2,3 and structure (the pair direct correlation function in Fig.4) , that can be achieved by implementing the sum rules. The application to an inhomogeneous case in  Fig.~\ref{fig:sp-cav} is one example of where our investigation might head.Tackling situations of pronounced confinement is an important testing ground. Of course, achieving stability and accuracy in spherical confinement does not guarantee similar performance in other confining geometries. This topic is one to be investigated further, accompanied by comparison with detailed simulation results.}

So far our results have been restricted to a one-component HS system. As laid out in \cite{Melih2024} the test-particle sum rules apply to any pair potential. The two sum rules could find applications in testing the accuracy of new density functionals for a wide variety of model fluids, especially those that include attractive contributions and those based on machine learning approaches. More generally, one might also consider binary mixtures. The sum rules can be generalized to this case. An obvious extension is to HS mixtures. It would be interesting to study the structure and thermodynamics of a binary HS mixture using the optimal parameters determined for the one-component HS fluid and consider whether these parameters remain optimal in the binary system. Note that the tensorial weight functions for a binary mixture might require some additional fine-tuning in order to account for the size asymmetry, as was reported in \cite{Cuesta2002}.

\appendix
\section{Thermodynamic Quantities}\label{app::thermoQuant}

From the extended Lutsko functionals, Eq.\eqref{eq::lutsko-wb} and Eq.\eqref{eq::lutsko-wbII}, the excess chemical potential $\mu_{ex}$ and the reduced isothermal compressibility $\chi_T$ are easily obtained from the bulk free energy density $\Phi(\rho)$.

We find
\begin{align}
    \beta\mu_\text{ex}^{\text{LK-WB}}(A,B) &= \beta\mu^{\text{CS}}_\text{ex}+(8A+2B-9)\frac{\eta(1+2\eta-\eta^2)+(1-\eta)^3\log(1-\eta)}{9(1-\eta)^3}\\\nonumber
    \chi_T^{\text{LK-WB}}(A,B) &=\chi_T^{\text{CS}}\left(1+(8A+2B-9)\frac{\eta^2}{9}\frac{9-4\eta+\eta^2}{1+4\eta+4\eta^2-4\eta^3+\eta^4}\right)^{-1},
\end{align}
for the WB and
\begin{align}
    \beta\mu_\text{ex}^{\text{LK-WBII}}(A,B) &= \beta\mu^{\text{CS}}_\text{ex}-(8A+2B-9)\frac{\eta(1-7\eta+6\eta^2-2\eta^3)+(1-\eta)^3\log(1-\eta)}{9(1-\eta)^3}\\\nonumber
    \chi_T^{\text{LK-WBII}}(A,B) &=\chi_T^{\text{CS}}\left(1+(8A+2B-9)\frac{\eta^2}{9}\frac{9-8\eta+7\eta^2-2\eta^3}{1+4\eta+4\eta^2-4\eta^3+\eta^4}\right)^{-1},
\end{align}
for the WBII version of the Lutsko functional, where $\beta\mu^{\text{CS}}$ and $\chi_T^{\text{CS}}$ refer to the corresponding CS quantities :
\begin{equation}
    \beta\mu^\text{CS}_\text{ex} = \frac{\eta(8-9\eta+3\eta^2)}{(1-\eta)^3},\quad \chi_T^\text{CS} = \frac{(1-\eta)^4}{1+4\eta+4\eta^2-4\eta^3+\eta^4}. 
\end{equation}

In both cases the CS results are recovered for $8A+2B-9=0$. Clearly any choice with $8A+2B\neq 9$ ensures that $\mu_{ex}$ and $\chi_T$ deviate from the CS results at second order in the density.

\section{Virial Coefficients for the HS fluid}\label{app::Virials}
The HS virial coefficients $B_n$ follow from the density expansion of the equations of state, Eq.\eqref{eq::Lutsko-wb-eos} and Eq.\eqref{eq::Lutsko-wbII-eos}, where we introduce the quantity $C=\frac{1}{3}(8A+2B-9)$ for convenience.

\begin{table}[h]
    \centering
    \scalebox{1.35}{
    \begin{tabular}{|c|c|c|c|c|}
        \hline
        $n$ & $B^\text{exact}_n$ & $B^\text{LK}_n$ & $B^\text{LK-WB}_n$ & $B^\text{LK-WBII}_n$ \\
        \hline\hline
        2 & 4 & 4 & 4 & 4 \\
        3 & 10 & 10+$C$ & 10+$C$ & 10+$C$ \\
        4 & 18.36 & 19+3$C$ & 18+$\frac{8}{3}C$ & 18+$\frac{7}{3}C$ \\
        5 & 28.22 & 31+6$C$ & 28+5$C$ & 28+$\frac{13}{3}C$ \\
        6 & 39.82 & 46+10$C$ & 40+8$C$ & 40+7$C$ \\
        7 & 53.34 & 64+15$C$ & 54+$\frac{35}{3}C$ & 54+$\frac{31}{3}C$ \\
        \hline
    \end{tabular}
    }
    \caption{Virial coefficients $B_n$ from the Lutsko functionals together with the exact values $B^{\text{exact}}_n$.}
    \label{tab:virials}
\end{table}

We infer from Table.\ref{tab:virials} that the predictions from the Lutsko functionals agree with each other for $n=3$ but already differ from the exact value if $C\neq0$. The higher virial coefficients differ between the various Lutsko functionals.

\section{Pair Direct Correlation Function}\label{app::DCF}
The pair direct correlation function $c^{(2)}(r)$ of the bulk liquid , with packing fraction $\eta$ , is obtained through
\begin{equation}\label{eq::c2-def}
    c^{(2)}(\mathbf{r}-\mathbf{r}')=-\sum_{\alpha,\beta}\frac{\partial^2\Phi}{\partial n_\alpha \partial n_\beta}\,\omega_\alpha\otimes\omega_\beta(\mathbf{r}-\mathbf{r}'),
\end{equation}
where $\omega_\alpha\otimes\omega_\beta$ is defined as
\begin{equation}\label{eq::c2-def-2}
    \omega_\alpha\otimes\omega_\beta(\mathbf{r}-\mathbf{r}')\equiv \int d\mathbf{r}''\,\omega_\alpha(\mathbf{r}-\mathbf{r}'')\omega_\beta(\mathbf{r}'-\mathbf{r}'').
\end{equation}
These convolutions can be further simplified by making use of spherical symmetry, i.e. $c^{(2)}$ solely depending on $r$.

Examining the various combinations of second order derivatives gives in the LK-WB case the following expression
\begin{equation}\label{eq::c2-lkwb}
    c^{(2),\text{LK-WB}}(r)=\left[a_0(\eta)+a_1(\eta)\frac{r}{\sigma}+a_3(\eta)\left(\frac{r}{\sigma}\right)^3\right]\Theta(\sigma-r),
\end{equation}
where $\Theta $ is the Heaviside function  and the coefficients read :
\begin{align}\label{eq::c2-lkwb-coeffs}
	a_0(\eta) &= \frac{-9-36\eta+(45-64A-16B)\eta^2+(16A+4B)\eta^3}{9(1-\eta)^4}\\\nonumber
	a_1(\eta) &= \frac{1}{18}\left(\frac{24A\log(1-\eta)}{\eta}-\frac{3 N_1(\eta)}{(1-\eta)^4}\right)\\\nonumber
	a_3(\eta) &= \frac{1}{18}\left(-\frac{36A\log(1-\eta)}{\eta}+\frac{N_2(\eta)}{(1-\eta)^4}\right).
\end{align}
and where for convenience we have introduced the polynomials of third degree :
\begin{align}\label{eq::n1-n2}
    N_1(\eta) &= -8A+(-45+40A+6B)\eta+(36-104A-24B)\eta^2+(9+24A+6B)\eta^3\\\nonumber
    N_2(\eta) &= -36A+(-9+144A+18B)\eta+(-36-196A-40B)\eta^2+(45+40A+10B)\eta^3
\end{align}

Similarly, the expression for the LK-WBII case is
\begin{equation}\label{eq::c2-lkwb2}
    c^{(2),\text{LK-WBII}}(r)=\left[b_0(\eta)+b_1(\eta)\frac{r}{\sigma}+b_3(\eta)\left(\frac{r}{\sigma}\right)^3\right]\Theta(\sigma-r)
\end{equation}
with
\begin{align}\label{eq::c2-lkwb2-coeffs}
    b_0(\eta) &= -\frac{9+36\eta+(-27+56A+14B)\eta^2-2(9+8A+2B)\eta^3+2(4A+B)\eta^4}{9(1-\eta)^4} \\\nonumber
    b_1(\eta) &= \frac{(6-4A)\log(1-\eta)}{3\eta}+\frac{N_3(\eta)}{6(1-\eta)^4}\\\nonumber
    b_3(\eta) &= \frac{2(A-1)\log(1-\eta)}{\eta}+\frac{N_4(\eta)}{18(1-\eta)^4},
\end{align}
where we have now defined the fourth degree polynomials:
\begin{equation}\label{eq::n3}
    N_3(\eta) =12-8A+(3+16A-6B)\eta+(32A+30+24B)\eta^2-5(9+2B)\eta^3+(8A+4B)\eta^4    
\end{equation}
and
\begin{gather}\label{eq::n4}
    N_4(\eta) = -36+36A+(-108A+117+18B)\eta+
    (112A-198-44B)\eta^2+\\\nonumber(117-92A+22B)\eta^3+(4A-8B)\eta^4 .
\end{gather}

\begin{acknowledgments}
We thank Jim Lutsko for valuable discussions at the start of this study. We acknowledge continuing support from the White-Bear Institute in Bristol that promotes fundamental research into the natural sciences.
\end{acknowledgments}

% \textbf{
%     [ Melih, The References do not have a consistent style. Do these need to  be inputted in the PRE style or does the submission process automatically format correctly?]}

% Create the reference section using BibTeX:
\bibliographystyle{apsrev}
\bibliography{lutsko_wb.bib}

\end{document}